\documentclass[useAMS,usenatbib,usegraphicx]{mn2e}

\title[SPB stars in NGC 371]{SPB stars in the open SMC cluster NGC 371}
\author[Karoff et al.]{
C. Karoff$^{1}$\thanks{E-mail: karoff@phys.au.dk}, T.  Arentoft$^{1}$, L. Glowienka$^{1}$, C. Coutures$^{2}$, T. B. Nielsen$^{1}$,  \and G. Dogan$^{1}$, F. Grundahl$^{1}$  \& H. Kjeldsen$^{1}$\\
$^{1}$Department of Physics and Astronomy, University of Aarhus, DK-8000 Aarhus C, Denmark\\
$^{2}$ Institut d'Astrophysique de Paris, CNRS, Universite Pierre et Marie Curie UMR7095,
 98bis  Boulevard Arago, 75014 Paris, France }
\begin{document}

%\date{2007}

\pagerange{\pageref{firstpage}--\pageref{lastpage}} \pubyear{2008}

\maketitle

\label{firstpage}

\begin{abstract}
Pulsation in $\beta$ Cep and SPB stars are driven by the $\kappa$ mechanism which depends critically on the metallicity. It has therefore been suggested that $\beta$ Cep and SPB stars should be rare in the Magellanic Clouds which have lower metallicities than the solar neighborhood. To test this prediction we have observed the open SMC cluster NGC 371 for 12 nights in order to search for $\beta$ Cep and SPB stars.  Surprisingly, we find 29 short-period B-type variables in the upper part of the main sequence, many of which are probably SPB stars. This result indicates that pulsation is still driven by the $\kappa$ mechanism even in low metallicity environments. All the identified variables have periods longer than the fundamental radial period which means that they cannot be $\beta$ Cep stars. Within an amplitude detection limit of 5 mmag no stars in the top of the HR-diagram show variability with periods shorter than the fundamental radial period. So if $\beta$ Cep stars are present in the cluster they oscillate with amplitudes below 5 mmag, which is significantly lower than the mean amplitude of $\beta$ Cep stars in the Galaxy.   We see evidence that multimode pulsation is more common in the upper part of the main sequence than in the lower. We have also identified 5 eclipsing binaries and 3 periodic pulsating Be stars in the cluster field.
\end{abstract}

\begin{keywords}
star: early-type --- stars: oscillations (including pulsations) --- galaxies: Magellanic Clouds
\end{keywords}

\section{Introduction}
$\beta$ Cep and slowly pulsating B stars (SPB) pulsate due to the $\kappa$-mechanism activated by the metal opacity bump \citep{1992ApJ...393..272C}. This give rise to $p$-mode pulsation in $\beta$ Cep stars and $g$-mode pulsation in SPB stars. However the theoretical standard models do not predict the presence of pulsation in $\beta$ Cep and SPB stars in low-metallicity environments such as the Magellanic Clouds.

Analysis of OGLE data have indeed shown that $\beta$ Cep and SPB stars exist in the Magellanic Clouds \citep{2002A&A...388...88P, 2004ASPC..310..225K, 2006MmSAI..77..336K} though it is not clear if the rate of $\beta$ Cep and SPB stars is lower than or equal to the rate in the Galaxy. We have therefore observed the open SMC cluster NGC 371 for 12 nights in order to search for $\beta$ Cep and SPB stars. Observing a single cluster instead of the the entire SMC has a number of advantages, -- e.g. the age and metallicity of the cluster can be constrained from fitting isochrones to the HR-diagram.

The metallicity of the SMC has been measured to between $Z$ = 0.001 and $Z$ = 0.004 \citep[see][and references therein]{1999A&A...346..459M}, while \citet{2007MNRAS.375L..21M, 2007CoAst.151...48M} have calculated instability domains of $\beta$ Cep and SPB stars using improved opacities and metal abundances which do not predict pulsation in $\beta$ Cep and SPB stars for $Z$ $<$ 0.005. We should therefore not expect to find any $\beta$ Cep or SPB stars in the SMC.

In order to test this prediction we have chosen to observe the cluster NGC 371 as this cluster has been observed before by \citet{1994IAUS..162...29K} who saw some signs of variability, though no clear evidence of pulsation was seen.

The observations presented in this paper have been made with a single telescope. This means that clear mode and frequency determination can not be obtained and the frequencies can not be used for modeling. Instead this work presents a number of candidate SPB stars for followup observations with at least dual-side photometry and with spectroscopy.

The paper is arranged as follows. Section~2 presents the cluster. Section~3 describes how we obtained, reduced and analyzed the data. The identified eclipsing binaries are presented in Section~4, the Be stars in Section~5 and the  pulsating stars in the upper part of the main sequence  in Section~6. A summary and concluding remarks are found in Section~7.
\section{The target cluster: NGC 371}
NGC 371 ($\alpha_{2000}, \delta_{2000}$ = 11$^{\mathrm{h}}$03$^{\mathrm{m}}$25$\fs$0,$-$72$^{\circ}$04$^{\prime}$40$\farcs$0) is a young open cluster in the SMC. \citet{2006ApJ...652..458W} estimated a log(age) of 6.7 based on isochrone fitting to OGLE data assuming a metallicity of $Z$ = 0.002. As there are no high-resolution spectra available for NGC 371 it has not been possible to estimate the metallicity autonomously.

The field of the cluster is shown in Fig.~1 with the identified variable stars marked.
\begin{figure}
\begin{center}
          \includegraphics[width=\columnwidth]{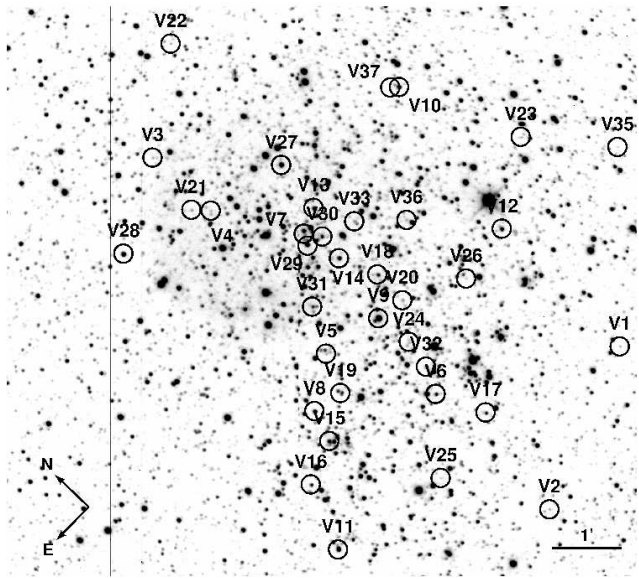}
\end{center}
\label{fig01}
\caption{DFOSC image of NGC371, with the variables marked.}
\end{figure}
\section{The data and the data reduction}
The observations were obtained with the DFOSC instrument at the Danish 1.54-m telescope at ESO, La Silla during 12 nights in August and September 2005. 763 frames were collected in $I$ and 623 in $B$. The same pointing was maintained during the observations, i.e., the stars were kept at fixed positions on the CCD (within a few pixels) during the observing run. Because of the crowding in the field the observations were always made in focus and the exposure times were then adjusted according to the seeing so that only the brightest 5 $\%$ of the stars were saturated. This resulted in exposure times of approximately 20 s in $I$ and 60 s in $B$. In Fig.~2 all the data are shown for one of the bright stars in NGC 371, illustrating the time distribution of the data set.

The CCD images were calibrated using standard procedures, i.e. a master BIAS was subtracted from each frame before the frames were divided by a master sky-flat in each filter. The master BIAS was obtained from a large number of frames from the whole observing run and the master sky-flats were obtained as the median of a large number of evening and morning sky-flats. We checked that both the BIAS and the flat-fields were indeed stable over the length of the observing run.

The photometric reductions were done using the software package MOMF \citep{1992PASP..104..413K}. MOMF applies a very robust algorithm combining PSF and aperture photometry in semi-crowded fields.

Some of the light curves showed residuals of systematic trends caused by, e.g., air mass, cloud cover, and temperature variations (these residuals are sometimes referred to as "red noise"). We therefore applied the algorithm introduced by \citet{2005MNRAS.356.1466T} in order to correct for systematic effects. This clearly improved the rms noise level, in particular in the brightest stars.

The resulting light curves had rms noise levels over the whole observing run ranging from a few mmag in the bright end and  up to 50 mmag in the faint end. Some of the light curves showed night-to-night drifts, which means that the rms noise level in the individual nights was lower than over the whole observing run. Fig.~2 shows the precision in a typical bright star. 

\begin{figure}
\begin{center}
          \includegraphics[width=\columnwidth]{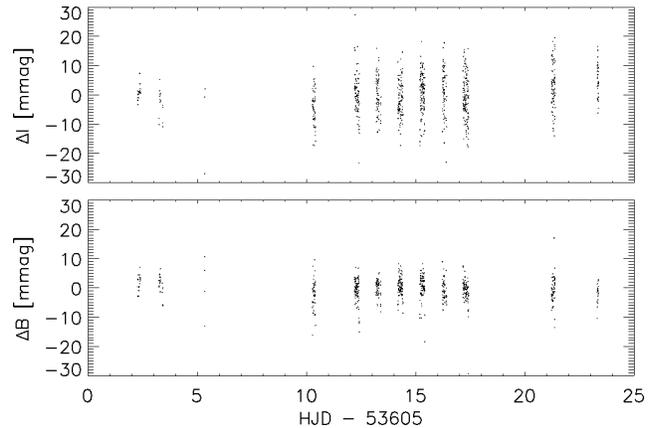}
\end{center}
\label{fig02}
\caption{Light curves of one of the bright stars in NGC 371 showing the precision level and data sampling.}
\end{figure}
\subsection{Analysis of differential light curves}
We obtained differential light curves of 6842 stars in the 13.5' $\times$ 13.5' FOV and we searched for variability in all these light curves. The light curves were kept in the instrumental system, but calibrated $UBVI$ photometry was obtained from \citet{2002AJ....123..855Z}.

We used two complementary algorithms to search for variability; the analysis of variance periodograms \citep{1996ApJ...460L.107S}  and simultaneous iterative sine wave fitting \citep{1995A&A...301..123F} based on the Lomb periodogram \citep{1976Ap&SS..39..447L}. We searched for variability both with and without using statistical weights \citep{1995A&A...301..123F}. The statistical weights were assigned as the standard deviation of all data points separated by less than 15 minutes from a given data point. For the analysis of variance approach we made visual examination of all light curves which had fit qualities better than 0.9 and for the Lomb periodogram approach we made visual examination of all light curves where the same peak in the periodogram was present in both $B$ and $I$ with a S/N higher than 4 in amplitude \citep{1993A&A...271..482B}. In total we ended up with a list of a little over a hundred stars that were selected for further analysis.

In the visual examination, stars were rejected mainly because of one of the following three causes: stars were placed outside the cluster region, stars were placed close to a hot pixel or a bad column, or the variability was not the same in the two filters. This gave us the list of the 37 stars presented in this paper. Light curves of all the variable stars will be added to CDS.

 It is not possible to determine if a star is a true cluster member or not, as we only have two dimensional images available. To determine if a star is likely to be a true cluster member in three dimension a spectroscopic analysis is needed. Therefore we can not give a reliable estimate of the number of cluster members and some of the identified variable stars could turn out not to be clusters members.
\section{Eclipsing binaries}
\begin{figure}
\begin{center}
          \includegraphics[width=\columnwidth]{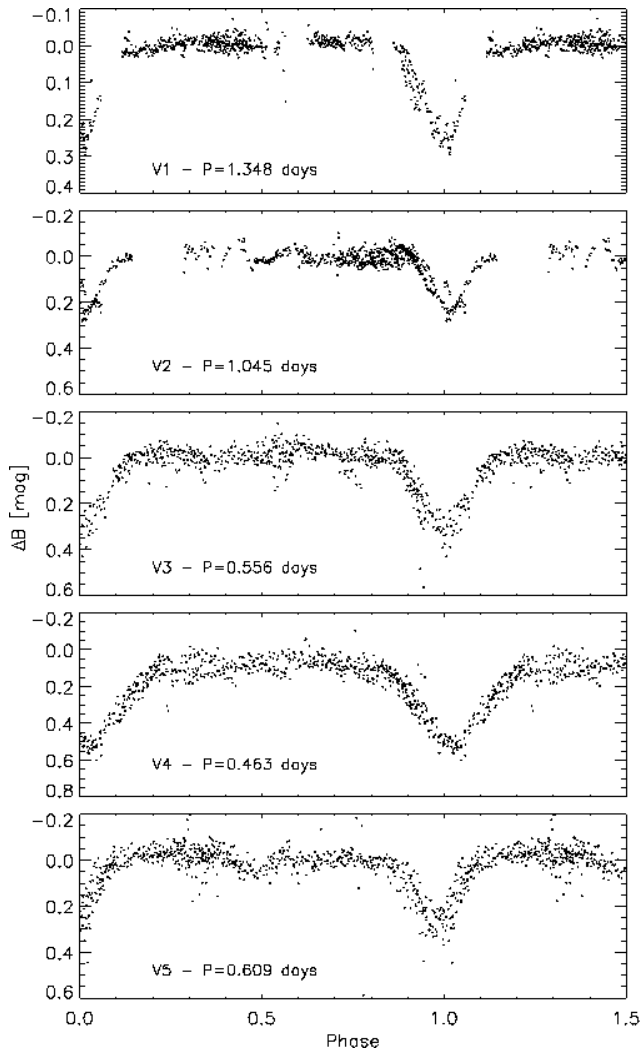}
\end{center}
\label{fig03}
\caption{Phase light curves of the 5 eclipsing binaries identified in NGC 371}
\end{figure}
We have detected 5 eclipsing binaries (phased light curves are shown in Fig.~3). V1, V2 and V3 are located on the edge of the cluster, which means that they might not be cluster members.  V4 and V5 are located safely inside the cluster and these stars could be important in determining the precise distance and age of the cluster.  This information will again be important when trying to model the excitation mechanisms of $\beta$ Cep and SPB stars based on data from this cluster.

V2 also shows some signs of pulsation with a frequency of 6.2 c/d, which suggest that this could be a binary system with one of the members being a pulsating star. But more photometry will be needed in order to evaluate the phenomenon properly, -- i.e removing the eclipse from the light curve before analyzing the pulsation.
\begin{table}
\caption{Stellar parameters for 5 eclipsing binaries in NGC 371. Periods are in days}
\centering
\begin{tabular}{lccccc}
\hline \hline
ID & $\alpha_{2000}$ & $\delta_{2000}$ & B & B -- I & P \\
\hline
V1 & 01 03 04.1 & -72 08 04 & 18.74 & 0.07 & 1.348 \\
V2 & 01 03 34.9 & -72 08 60 & 18.64 & 2.68 & 1.045 \\
V3 & 01 03 40.5 & -72 01 26 & 18.86 & -0.17 & 0.556 \\
V4 & 01 30 40.0 & -72 02 33 & 19.14 & -0.22 & 0.463 \\
V5 & 01 03 43.5 & -72 05 11 & 18.64 & -0.23 & 0.609 \\
\hline
\end{tabular}
\label{tab1}
\end{table}

\section {Be stars}
\begin{figure}
\begin{center}
          \includegraphics[width=\columnwidth]{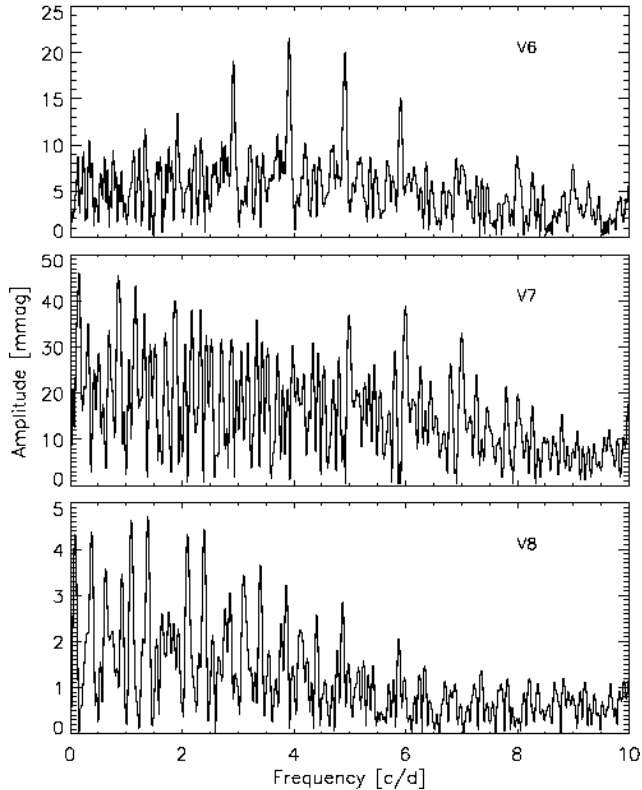}
\end{center}
\label{fig04}
\caption{Amplitude spectra for 3 Be stars in NGC371. Note that the scale on the y-axis is different for the three stars.}
\end{figure}
Three of the detected variable stars match with the stars identified as Be stars by \citet{2006ApJ...652..458W}. They have identified 118 Be stars and 11 candidate Be stars in NGC371 based on two-color diagrams of $B$, $V$, $R$ and $H \alpha$ photometry. Though the nature of the variability of Be stars is believed to be transient, Be stars might also exhibit $g$- or $p$-mode pulsation. It is therefore clear that these three stars can be used to gain understanding of the relation between $g$- and $p$-mode pulsation and the variability of Be stars in low-metallically environments. The amplitude spectra of the three Be stars are shown in Fig.~4. The light curves of the three Be stars all show coherent variability, which indicate that the variability originates from pulsation and not from transient events like activity.
\begin{table}
\caption{Stellar parameters for 3 Be stars in NGC 371. The id's in the second column (ID II) are from \citet{2006ApJ...652..458W}.	}
\centering
\begin{tabular}{llcccc}
\hline \hline
ID &  ID II & $\alpha_{2000}$ & $\delta_{2000}$ & B & B -- I\\
\hline
V6 & WBBe 43 & 01 03 34.5 & -72 06 42 & 16.65 & 0.35\\
V7 & WBBe 110 & 01 03 30.5 & -72 03 46 & 17.99 & -0.34\\
V8 & WBBe 18 & 01 03 52.6 & -72 05 39 & 15.73 & 0.21\\
\hline
\end{tabular}
\label{tab2}
\end{table}
\begin{figure}
\begin{center}
          \includegraphics[width=\columnwidth]{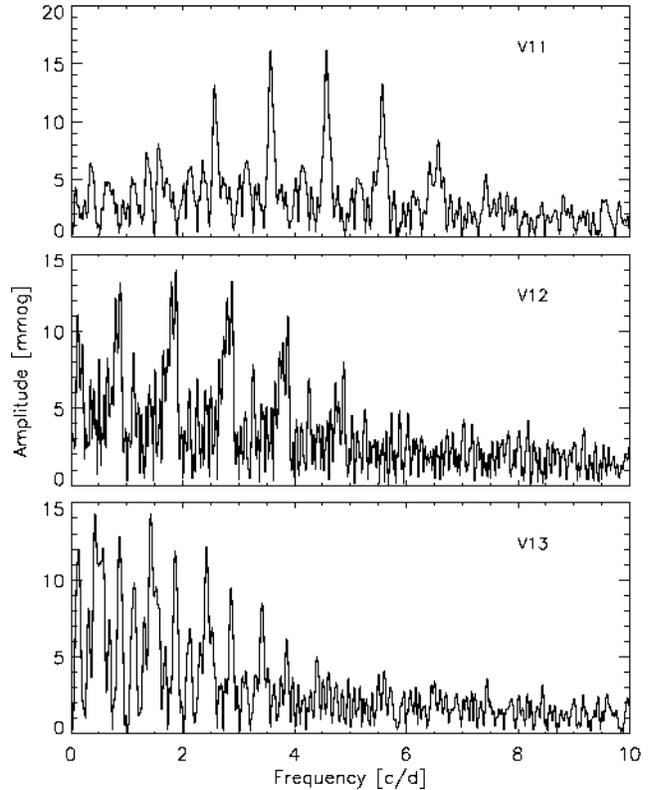}
\end{center}
\label{fig05}
\caption{Amplitude spectra for three pulsating stars in the upper part of the main sequence of NGC 371 }
\end{figure}

\begin{figure}
\begin{center}
          \includegraphics[width=\columnwidth]{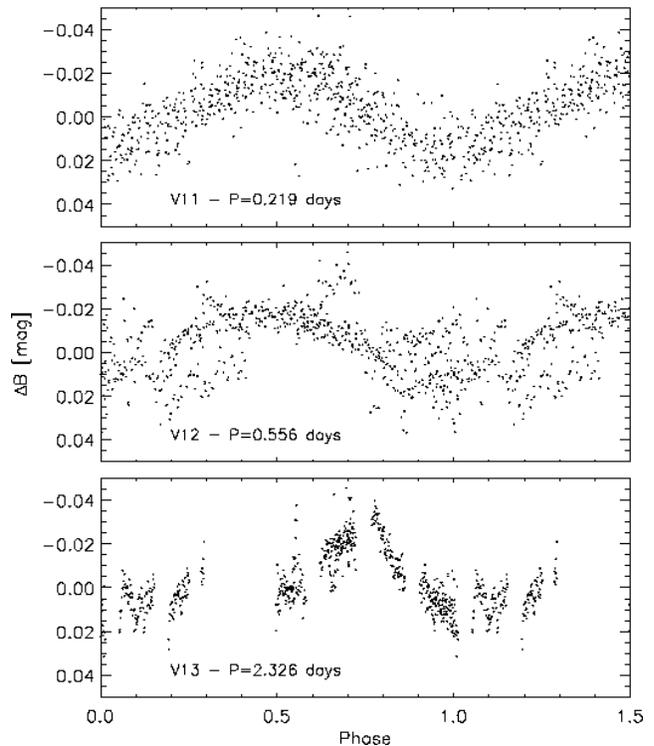}
\end{center}
\label{fig06}
\caption{Phased light curves for three pulsating stars in the upper part of the main sequence of NGC 371 }
\end{figure}
\section{Pulsating stars in the upper part of the main sequence}
The frequency analysis of the variable stars in the upper part of the main sequence  followed \citet{2007A&A...465..965A}. This means that we made a simultaneous least-squares fit to all the frequencies with a S/N higher than 4 in the amplitude spectra of both the $I$ and the $B$ filter data. For each frequency we manually inspected the amplitude spectra, in order to find the frequency value that best described the variation in both filters. This step could not be made completely objective as daily aliases were present in the amplitude spectra because the observations were made from a single site. Therefore some of the detected frequencies might be shifted 1--2 c/d from the true oscillation frequencies. The uncertainties on the amplitudes and phases were calculated based on the error matrix in the least-squares fitting procedure, though we also used Monte Carlo simulations to check the consistency of these uncertainties [this was done with the use of Period04 \citep{2005CoAst.146...53L}]. This procedure for calculating the uncertainties assumes that we know the frequencies, -- i.e. that we know which peak reflect the oscillation mode. The calculated uncertainties does therefore not reflect the uncertainty originating from mismatch between the true oscillation mode and the daily aliases. 

All the detected frequencies for the  pulsating  stars are given in Table~3 together with color, amplitude and phase information,  three examples of amplitude spectra of pulsating stars are shown in Fig.~5 and phased light curves are shown in Fig.~6.
\begin{table*}
\caption{Stellar parameters for the  pulsating  stars in NGC 371. Frequencies ($\nu$) are in c/d, amplitudes ($A_B$) are in mmag, phase differences ($\phi_I - \phi_B$) are in radians and pulsation constants $Q$ are in d$^{-1}$. The quoted errors on the amplitude ratios and phase differences are based on the error matrix in the least-squares fitting procedure.}
\centering
\begin{tabular}{lcccccccccc}
\hline \hline
ID     & $\alpha_{2000}$    & $\delta_{2000}$    & B  [mag]                   & B -- I                 & $\nu_i$          		& $\nu$              & $A_B$  		& $A_I/A_B$     & $\phi_I - \phi_B$  	& $Q$\\
\hline
V9	&	01 03 32.0		&	 -72 05 21		&	13.72	&      -0.58		&	$\nu_1$		&	1.95		&      6.89		&     0.46(7)	&     -0.73(3)		& 0.085	\\    
	&					&				&			&			&	$\nu_2$		&	2.12		&      4.64		&     0.25(10)	&     -0.34(4)		& 0.078   \\
V10	&	01 02 58.9		&	-72 03 14		&	15.57	&	-0.33		&	$\nu_1$		&       1.78		&      12.98	&     0.68(7)	&      0.19(2)		& 0.205	\\   
	&					&				&			&			&	$\nu_2$		&	1.70		&      5.78		&     1.11(19)	&     -0.24(2)		& 0.215	\\    
V11	&	01 04 37.5		&	-72 09 32		&	17.75	&	-0.33		&	$\nu_1$		&      4.57		&      16.70	&     0.52(11)	&     -0.03(3)		& 0.137	\\   
V12	&	01 03 04.2		&	-72 05 42		&	15.89	&	-0.47		&	$\nu_1$		&      1.80		&      12.97	&     1.00(9)	&	0.09(1)		& 0.229	\\    
V13	&	01 03 25.9		&	-72 03 36		&	15.79	&      -0.52		&	$\nu_1$		&	0.43		&      11.94	&     1.76(10)	&	 0.07(1)		& 0.958	\\   
	&					&				&			&			&	$\nu_2$		&	0.87		&      9.17		&     1.57(11)	&       -0.03(2)		& 0.474   \\
V14	&	01 03 29.2		&	-72 04 21		&	15.95	&	-0.54		&	$\nu_1$		&       2.85		&      19.60	&     0.58(10)	&	-0.11(2)		& 0.145	\\    
	&					&				&			&			&	$\nu_2$		&	2.40		&      14.85	&     1.58(14)	&	-0.04(2)		& 0.172	\\    
V15	&	01 03 54.7		&	-72 06 05		&	16.24	&	-0.43		&	$\nu_1$		&       3.86		&      20.99         &	0.57(8)	&       -0.01(2)		& 0.123	\\    
	&					&				&			&			&	$\nu_2$		&	1.80		&      8.65		&     0.69(20)	&        0.03(4)		& 0.264   \\
V16  &	01 04 02.7		&	-72 06 21		&	16.70	&	-0.31		&	$\nu_1$		&       3.33		&      30.44	&      0.43(4)	&	  0.00(2)		& 0.151	\\   
	&					&				&			&			&	$\nu_2$		&       0.48		&      14.86	&     0.54(13)	&	 -0.04(4)		& 1.047	\\    
V17	&	01 03 30.4		&	-72 07 23		&	16.91      	&			&	$\nu_1$		&	 3.61		&      27.71	&      0.97(6)	&        -0.02(1)		& 0.152	\\   
	&					&				&			&			&	$\nu_2$		&	2.73		&      23.52	&      0.73(6)	&	  0.02(1)		& 0.202	\\    
V18	&	01 03 16.3		&	-72 04 55		&	16.66	&	-0.43		&	$\nu_1$		&       2.18		&      32.17	&      0.81(6)	&	  0.03(10)		& 0.230	\\     
V19	&	01 03 46.9		&	-72 05 44		&	17.72	&	-0.43		&	$\nu_1$		&      1.52		&      63.71	&      0.87(5)	&    	-0.01(4)		& 0.412	\\    
V20	&	01 03 26.5		&	-72 05 25		&	18.23	&	-0.29		&	$\nu_1$		&      1.94		&      73.13	&      0.97(4)	&        0.03(1)		& 0.382	\\   
V21	&	01 03 42.4		&	-72 02 21		&	18.43	&	-0.16		&	$\nu_1$		&      1.40		&      35.31	&      0.90(17)	&        0.02(3)		& 0.534	\\    
V22	&	01 03 23.2		&	-72 00 28		&	18.51	&	-0.32		&	$\nu_1$		&      0.52		&      44.84	&      0.72(11)	&	 -0.15(1)		& 1.438	\\   
V23	&	01 02 49.5		&	-72 04 59		&	18.92	&	-0.37		&	$\nu_1$		&	2.47		&      48.88	&      0.93(22)	&        0.00(4)		& 0.333	\\    
V24	&	01 03 31.2		&	-72 05 54		&	18.70	&	-0.37		&	$\nu_1$		&       0.74      	&	75.45     	&	1.04(6)	&    	-0.01(1)		& 1.051	\\    
V25	&	01 03 44.9		&	-72 07 35		&	19.06	&      	0.03		&	$\nu_1$		&	1.35		&      281.28	&      0.63(3)	&    	0.02(1)		& 0.610	\\   
V26	&	01 03 15.3		&	-72 05 50		&	19.01	&	-0.24		&	$\nu_1$		&       0.63		&      183.62	&      0.47(7)	&      -0.03(2)		& 1.307	\\   
V27	&	01 03 24.7		&	-72 02 49		&	14.99	&	 -0.53	&	$\nu_1$		&	2.52		&      5.09		&     1.20(16)	&	 0.02(2)		& 0.133	\\   
V28	&	01 03 57.0		&	-72 02 07		&	15.27	&	-0.45		&	$\nu_1$		&       3.03		&      10.82	&     1.10(12)	&	 -0.05(2)		& 0.121	\\   
V29	&	01 03 31.6		&	-72 03 54		&	15.32	&	-0.56		&	$\nu_1$		&      1.69		&      15.39	&      0.62(14)	&      -0.01(3)		& 0.216	\\
	&					&				&			&			&	$\nu_2$		&	2.04		&      8.54		&     0.79(25)	&	 -0.13(6)		& 0.179	\\
V30	&	01 03 28.6		&	-72 03 58		&	16.59	&	-0.41		&	$\nu_1$		&      1.74		&      19.77	&	1.13(16)	&      0.02(3)		& 0.288	\\   
V31	&	01 03 39.0		&	-72 04 34		&	16.90	&	-0.46		&	$\nu_1$		&      1.30		&      16.49	&      1.79(24)	&	-0.08(2)		& 0.423	\\   
V32	&	01 03 32.2		&	-72 06 19		&	17.28	&	-0.18		&	$\nu_1$		&     	1.64		&      29.46	&      1.22(10)	&     0.04(1)		& 0.336	\\    
V33	&	01 03 22.3		&	-72 04 08		&	18.27	&	-0.09		&	$\nu_1$		&      0.79		&      23.94	&      1.10(12)	&     -0.14(2)		& 0.939	\\    
V34	&	01 02 36.6		&	-72 02 33		&	18.97	&	-0.01		&	$\nu_1$		&      1.33		&      72.25	&      1.58(17)	&     -0.01(2)		& 0.619	\\    
V35	&	01 02 38.2		&	-72 06 02		&	19.21	&	0.11		&	$\nu_1$		&       0.38		&      170.37	&      0.61(3)	&    -0.09(1)		& 2.308	\\   
V36	&	01 03 15.2		&	-72 04 39		&	19.01	&	-0.1		&	$\nu_1$		&      1.24		&      164.74	&      0.83(6)	&     -0.01(1)		& 0.664	\\    
	&					&				&			&			&	$\nu_2$		&	0.50		&      134.43	&      0.56(12)	&     -0.04(5)		& 1.647	\\    
V37	&	01 03 00.1		&	-72 06 58		&	19.12	&	-0.32		&	$\nu_1$		&       0.73		&      250.32      &	0.60(5)	&    -0.03(1)		& 1.202	\\    
\hline
\end{tabular}
\label{tab3}
\end{table*}
\subsection{Color-magnitude-diagram of NGC 371}
We have used the photometry from \citet{2002AJ....123..855Z} to make a $B$ versus $B-I$ color-magnitude-diagram of NGC 371 as shown in Fig.~7. The center of the cluster was placed on $\alpha_{2000}, \delta_{2000}$ = 11$^{\mathrm{h}}$03$^{\mathrm{m}}$25$\fs$0,$-$72$^{\circ}$04$^{\prime}$40$\farcs$0 and the size was selected as a circle with a radius of 5'. The color-magnitude-diagram shows the upper part of the main sequence, which is almost vertical, and the giant branch. All the identified variable stars which are not eclipsing binaries or Be stars are located on the upper part of the main sequence, which is also where the instability strips for $\beta$ Cep and SPB stars would be located if the metallicity of the cluster had been larger.  Instability analysis of models of hot main sequence stars of solar metallicity show a group of mostly p-mode pulsators where the longest period is the fundamental radial mode and a
second group of cooler stars with mostly g-modes and longer periods.  The first group can be identified with the classical $\beta$~Cep stars (early B-type stars) while the second group can be identified with the SPB stars (mid B-type stars).  Models based on revised solar metal abundances and different opacities show that the hot boundary of the SPB instability strip overlaps with the $\beta$ Cep instability strip  \citep{2007MNRAS.375L..21M, 2007CoAst.151...48M}.  At the metallicity of NGC 371, however, no pulsations are excited in the models as already mentioned. Therefore theory cannot be used as a guide in classification.  We will come back to this in the next section.

We have used the Padova evolutionary code described in \citet{2007arXiv0711.4922M} to calculate an isochrone for the color-magnitude-diagram of the cluster. The model used solar heavy elements mixture. We adopted a metalicity of $Z$=0.002  and a log(age) of 6.7 from \citet{2006ApJ...652..458W}. The distance modulus was chosen to 18.7 from \citet{2001AJ....122..220C} and the reddening to  $E(B-V)=0.085$ \citep{2000A&A...364..455L}. 

The isochrone is plotted on top of the color-magnitude-diagram in Fig.~7. Here it is seen that the isochrone generally fits the data nicely.  
\begin{figure}
\begin{center}
          \includegraphics[width=\columnwidth]{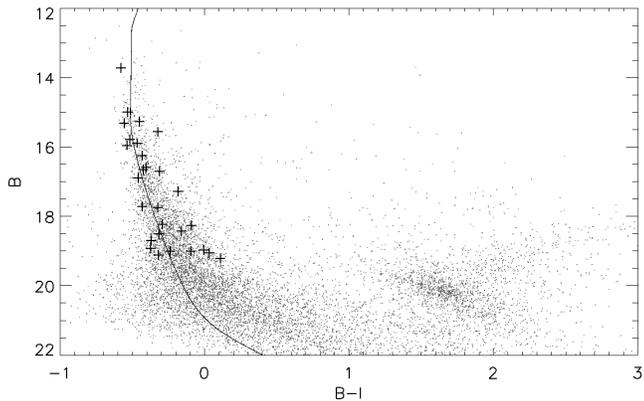}
\end{center}
\label{fig06}
\caption{Color-magnitude-diagram of NGC 371 with the identified  SPB candidates  marked. The solid line shows a isochrone fit to the observations.}
\end{figure}
\subsection{Period distribution in NGC 371}

In the Galaxy, $\beta$ Cephei stars have periods which are consistent with mostly p-mode pulsations.  The longest period is that of the fundamental
radial mode for which the pulsation constant $Q \approx 0.033$ d$^{-1}$ \citep{2005ApJS..158..193S}.  To classify stars in NGC 371 we need an estimate
of Q for each star:
\begin{eqnarray*}
\log Q = \log P + 1/2\log (M/M_\odot) - 3/4\log(L/L_\odot)&\\ + 3 \log (T/T_\odot).&
\end{eqnarray*}
where $P$ is the period in days, $M$ the mass, $L$ the luminosity and $T$ the effective temperature.  Since we do not have data to estimate the masses, luminosities and effective temperatures for individual stars, we use these values of the isochrone shown in Fig.~7.   The result is shown in Fig. 8
together with all the identified periods as a function of $B$ magnitude.  There the mode with the longest period in each star is plotted with large crosses, while the secondary modes with shorter periods are plotted with small diamonds.

In Fig.~8 it is seen that all stars have periods that are longer than the fundamental radial periods. We therefore identify all the stars as candidate SPB stars. Classification of the stars as bona fide SPB stars would require more time series observations to ensure that the periods are correct and exclude other sources of variability, such as close binaries.

If there are $\beta$ Cep stars in the cluster then we can assume two things about them; That they are generally brighter than the SPB stars and that they have periods shorter than the fundamental radial period. We do not detect any pulsating stars in this domain than fulfill the detection criteria given in Section  3.1 and therefore we do not detect any candidate $\beta$ Cep stars in NGC 371. 

In order to constrain an upper detection limit of $\beta$ Cep stars in the cluster we measured the highest peak in the amplitude spectrum between 8 and 20 c/d for all the stars brighter than 15.8 in $B$. A frequency range from 8 to 20 c/d is in agreement with the frequency range in which the $\beta$ Cep stars are expected for this cluster (see Fig.~8). For the brightest stars we could also have included periods up to 0.4 day, but this gave problems with the $1/f$ noise in the amplitude spectra of some of the stars. 

In Fig.~9 we have plotted the amplitude of the highest peaks as a function of magnitude together with the mean $V$ oscillation amplitude of galactic $\beta$ Cep stars from \citet{2005ApJS..158..193S}. It is seen that our detection limit is significantly lower than the mean $V$ oscillation amplitude of galactic $\beta$ Cep. If $\beta$ Cep stars are present in the cluster they have oscillation amplitudes significantly lower than the galactic $\beta$ Cep stars.

Though we do not detect any candidate $\beta$ Cep stars some of the stars do show variability between 8 and 20 c/d, but not with a S/N higher than 4 in both filters. In Fig.~9 we have marked the peaks which have S/N higher than 3.5 in $B$ and in Fig.~10 we have plotted the amplitude spectrum of one of them. The detection criteria of S/N higher than 4 in both filters is generally conservative and for longer regularly sampled data sets it is too high as the observations in the two filters are independent. But for the present data set it seems as a good indication of pulsation. More photometric data is needed in order to classify these stars as candidate $\beta$ Cep stars. On the other hand if these stars are $\beta$ Cep stars they would be very interesting targets as they would have shorter periods and lower oscillation amplitudes compared to the galactic $\beta$ Cep stars.

Though periods below 0.1 day are not common in $\beta$ Cep stars they have been seen in the very young cluster NGC 6231 \citep{1983MNRAS.205..309B, 2001A&A...380..599A}.

\begin{figure}
\begin{center}
          \includegraphics[width=\columnwidth]{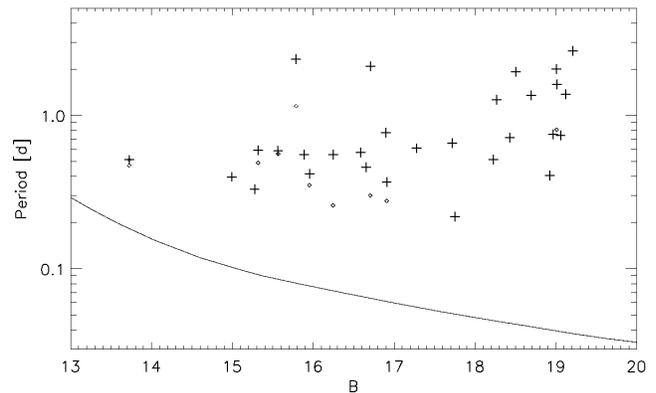}
\end{center}
\label{fig07}
\caption{Magnitude-period relation for the candidate SPB stars stars together with the fundamental radial period from the isochrone.  The mode with the longest period in each star is plotted with large crosses, while the secondary modes with shorter periods is plotted with small diamonds.  }
\end{figure}

\begin{figure}
\begin{center}
          \includegraphics[width=\columnwidth]{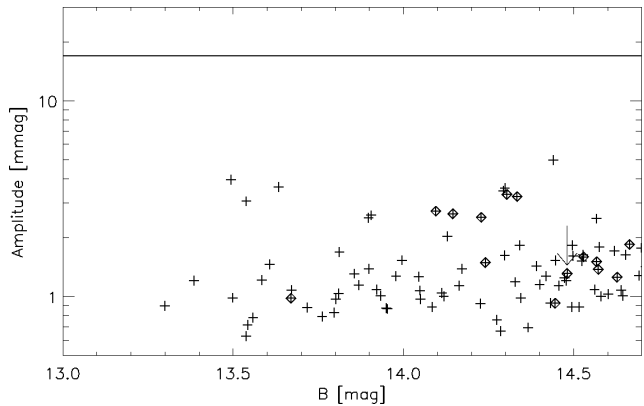}
\end{center}
\label{fig09}
\caption{Amplitudes of the highest peak as a function of magnitude for the stars brighter than the identified candidate SPB stars. The diamonds mark the peaks with a S/N higher than 3.5. The horizontal line at 17 mmag shows the mean oscillation amplitude of $\beta$ Cep stars in the Galaxy in$V$. The arrow marks the star plotted in Fig.~10.}
\end{figure}

\subsection{Amplitude ratio and phase difference}
We have also calculated and plotted the amplitude ratios and phase difference between the $B$ and $I$ filters (Fig.~11) in order to identify any systematic effects that could be compared with theoretical models as in \citet{2007A&A...465..965A} or used for mode identification \citep{1994A&AS..105..447H}. Though we do not find any systematic effects in phase difference versus amplitude ratio within the uncertainties (Fig.~11) we do see that the main part of the observed frequencies cluster around an amplitude ratio between 0.5 and 1.0 which is expected for oscillations in $\beta$ Cep stars with degrees between 0 and 2 \citep{1994A&AS..105..447H}.

 The $I/B$ amplitude ratio can also be used to exclude other causes of the variability than pulsation (e.g. ellipsoidal variables) as the $I/B$ amplitude ratios are general below unity \citep[see, e.g.,][and references therein]{2007A&A...465..965A}.

\subsection{Multimode pulsation}

The last thing to note about the frequencies is that all the multimode pulsators are found among the pulsating stars in the upper part of the main sequence (except for one). One reason for this could be that these stars are brighter, which in general gives a higher S/N, but some of the frequencies that we detect in the stars in the lower part of the main sequence have such a high amplitude that multimode pulsation should have been discovered if it was present (see Table~3).  

\section{Summary and Conclusions}
 We have identified 5 eclipsing binaries, 3 periodic Be stars and 29 candidate SPB stars in NGC371. 

The results indicates that excitation of oscillations in SPB stars is more common in low metallicity environments such as the SMC than predicted by standard stellar models. As a possible extension to the standard stellar models to account for this discrepancy could be to include local iron enhancement by diffusion and radiative accelerations \citep{2007CoAst.150..209M}.

If there are $\beta$ Cep stars in the cluster they would generally be brighter than the SPB stars and have periods shorter than the fundamental radial period and we do not detect pulsation in any stars in this domain with an upper limit on the oscillation amplitudes of 5 mmag. 

Though it is not possible to calculate a reliable number for the fraction of SPB stars in NGC 371 the absolute number of pulsating stars seems to be high compared to the fraction obtained for the Galaxy by \citet{2005ApJS..158..193S}. 

We see evidence that multimode pulsation is more common in upper part of the main sequence than in the lower.

We have also identified periodic pulsation in 3 stars that were previously identified as Be stars. These stars can used to study the relation between the Be phenomenon and $\beta$ Cep and SPB stars in low-metallicity environments. 

The results presented here strongly confirm and increase our interest in NGC 371 as an excellent laboratory for $\beta$ Cep and SPB stars.
\begin{figure}
\begin{center}
          \includegraphics[width=\columnwidth]{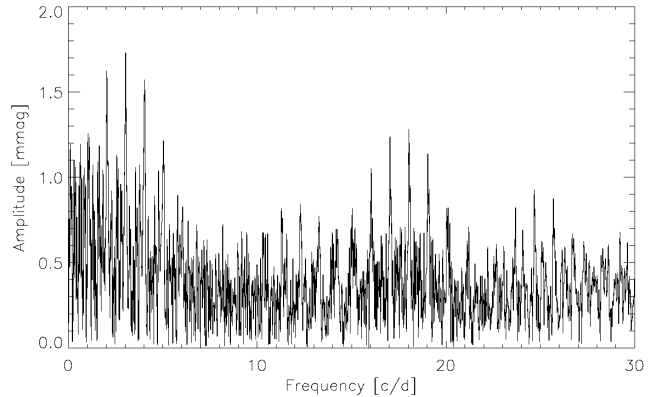}
\end{center}
\label{fig10}
\caption{Amplitude spectrum in $B$ of one of the stars showing signs of $\beta$ Cep variability between 8 and 20 c/d with a S/N below 4.}
\end{figure}
\section*{Acknowledgments}
 We thank the referee Luis Balona for suggesting that we examined the fundamental radial periods of the stars, which helped to improve the paper significantly. We also thank J{\o}rgen Christensen-Dalsgaard for useful suggestions.  The Danish Natural Science Research Council and the Instrument Center for Danish Astrophysics (IDA) are acknowledged for financial support. C.K. and F.G. acknowledges financial support from IDA. C.K, T.A., G.D.  and F.G. also acknowledges support from the Danish AsteroSeismology Centre.

\begin{figure}
\begin{center}
          \includegraphics[width=\columnwidth]{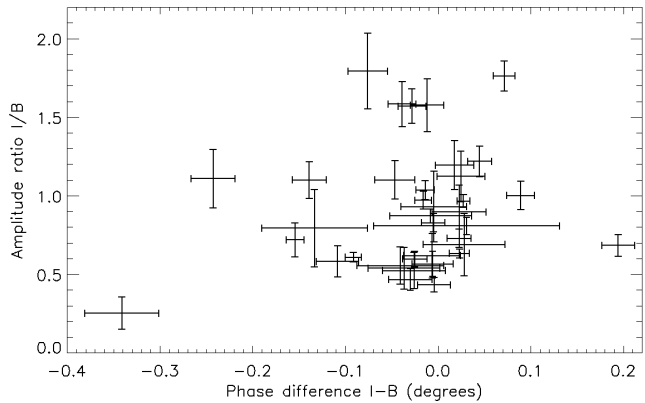}
\end{center}
\label{fig11}
\caption{Phase difference $I - B$ versus amplitude ratio $I/B$ for the candidate SPB stars}
\end{figure}

\label{lastpage}

\end{document}